\documentclass[aps,prl,floatfix,twocolumn,tightenlines,amsmath,amssymb]{revtex4}

\usepackage{graphicx}
\usepackage{amsmath}
\usepackage{amssymb}
\usepackage{epstopdf}
\usepackage{color}
\usepackage{tikz}

\newcommand{\ket}[1] {|#1 \rangle}

\newcommand*{\rom}[1]{\expandafter\@slowromancap\romannumeral #1@}
\newcommand*\circled[1]{\tikz[baseline=(char.base)]{
            \node[shape=circle,draw,inner sep=1pt] (char) {#1};}}

\begin{document}

\title{High Fidelity Single-Shot Singlet-Triplet Readout of Precision Placed Donors in Silicon}
\author{M. A. Broome, T. F. Watson, D. Keith, S. K. Gorman, M. G. House, J. G. Keizer, S. J. Hile, W. Baker, M. Y. Simmons}
\affiliation{Centre of Excellence for Quantum Computation and Communication Technology, School of Physics, University of New South Wales, Sydney, New South Wales 2052, Australia}
\date{\today}

\begin{abstract}
In this work we perform direct single-shot readout of the singlet-triplet states in exchange coupled electrons confined to precision placed donor atoms in silicon. Our method takes advantage of the large energy splitting given by the Pauli-spin blockaded (2,0) triplet states, from which we can achieve a single-shot readout fidelity of 98.4$\pm$0.2\%. We measure the triplet-minus relaxation time to be of the order 3~s at 2.5~T and observe its predicted decrease as a function of magnetic field, reaching 0.5~s at 1~T.
\end{abstract}

\maketitle

An increased ability to control and manipulate quantum systems is driving the field of quantum computation forward~\cite{Nielsen:2011,Ladd:2010vn,RevModPhys.86.153,RevModPhys.87.307}. The spin of a single electron in the solid-state has long been utilised in this context~\cite{petta2004,johnson2005triplet,hanson2007,pla2012,morello2010,buch2013,watson2014}, providing a superbly clean quantum system with two orthogonal quantum states that can be measured with over $99\%$ fidelity~\cite{watson2015}. As a natural next step, the coupling of two electrons at separate sites has been studied in gate-defined quantum dots~\cite{petta2004,koppens2007,maune2012}, as well as in donor systems~\cite{watson2015b,House:2015rz,dehollain2014}. In addition to being the eigenstates for two coupled spins, the singlet-triplet (ST) states of two electrons can form a qubit subspace, and have previously been utilised for quantum information processing~\cite{johnson2005triplet,petta2005,taylor2007,Prance2012,laird2010,Kim:2015jt}. Unlike in gate-defined quantum dots, donor systems do not require electrodes to confine electrons. The resulting decrease in physical complexity makes donor nano-devices very appealing for scaling up to many electron sites~\cite{watson2015b}.

\begin{figure*}
\begin{center}
\includegraphics[width=0.8\textwidth]{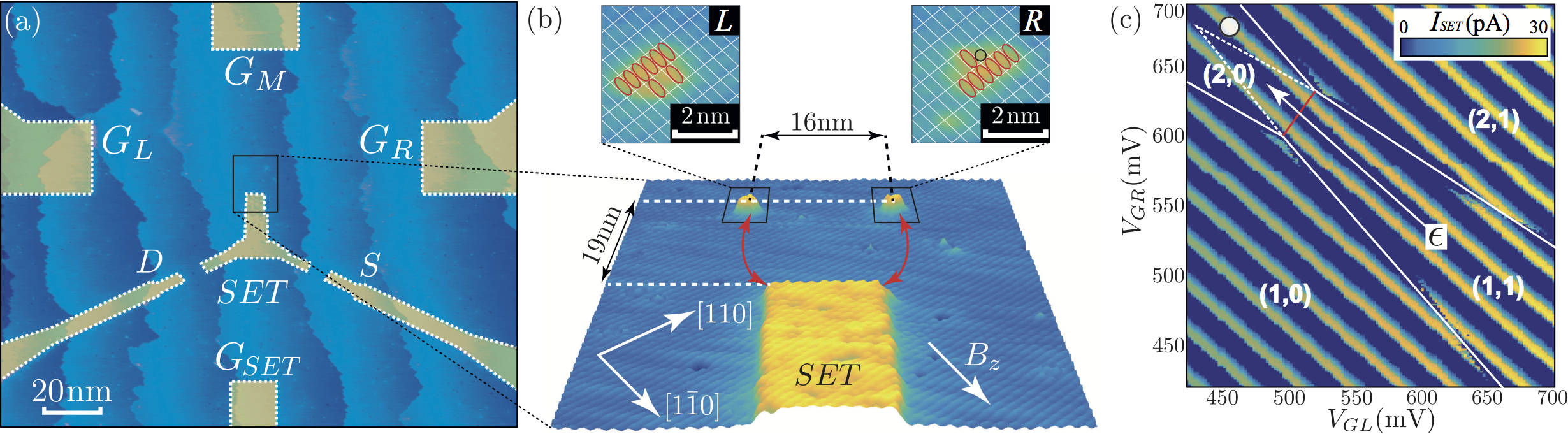}
\end{center}
\vspace{-0.5cm}
\caption{{\bf The (1,1)$\leftrightarrow$(2,0) charge transition in a double donor-dot.} a) A scanning tunnelling micrograph of the pre-dosed device showing the hydrogen resist (blue region) and silicon beneath (yellow overlay). Three gates $G_L$, $G_M$ and $G_R$ control the electrostatic environment of the donors-dots. The SET is tunnel coupled to source ($S$) and drain ($D$) and controlled predominantly by $G_{SET}$. b) Donor sites $L$ and $R$ are separated by $16{\pm}1$~nm, and are equidistant at $19{\pm}1$~nm from a single-electron-transistor (SET) charge sensor which also serves as an electron reservoir (red arrows). Insets are close-up images of $L$ and $R$ showing lithographic patches large enough for 2 and 1 P atoms respectively. c) A charge stability map showing the current through the SET as a function of voltages $\{V_{GL},V_{GR}\}$ near the (1,1)-(2,0)  transition. Current peaks running at $\sim45^\circ$ show Coulomb blockade of the SET and breaks in these lines correspond to single electron transitions of $L$ and $R$. The solid white lines indicate the dot-SET ground-state transitions, whereas the area enclosed by the dashed white lines shows the region where S(2,0) is the ground state and all T(1,1) states are metastable. The detuning axis $\epsilon$ is shown by the white arrow. Singlet-triplet readout is performed at the point shown by the circle marker.}
\label{fig:device}
\end{figure*}

In the (1,1) charge configuration the ST states are eigenstates if the exchange coupling is greater than any \textit{difference} in Zeeman energy between the two spins. The singlet and three triplet states are split only by the Zeeman energy in the cases of $\ket{T^+}{=}\ket{{\uparrow\uparrow}}$ and $\ket{T^-}{=}\ket{{\downarrow\downarrow}}$, and an exchange energy, $J$, for the singlet $\ket{S}{=}\left(\ket{{\uparrow\downarrow}}{-}\ket{{\downarrow\uparrow}}\right)/\sqrt{2}$ and $\ket{T^0}{=}\left(\ket{{\uparrow\downarrow}}{+}\ket{{\downarrow\uparrow}}\right)/\sqrt{2}$ states. However, in the (2,0) configuration all triplet states split from the singlet $\ket{S(2,0)}$ by a larger exchange interaction, $\Delta_{ST}$, measured in previous works to be ${>}5$meV for donors~\cite{weber2014}. The triplet states are therefore blocked from tunnelling from the (1,1)$\rightarrow$(2,0) charge configuration, known as Pauli spin-blockade.

Typically, direct ST readout is performed by charge discrimination between the (1,1) and (2,0) states below the ST energy splitting $\Delta_{ST}$. However, this relies on the charge sensor having a large enough differential capacitive coupling to each dot to discriminate between the two charge states. This is not possible in some architectures due to symmetry constraints, in particular for donors it is advantageous for multiple donor sites to be coupled equally to a charge sensor for independent readout and/or loading. The tightly confined electron wavefunction at each donor site therefore necessitates that they are equidistant from the charge sensor. As a consequence (1,1)${\leftrightarrow}$(2,0) charge transfer signals are often too small to detect directly in this architecture.

Until now single-shot readout of ST states in donors has been limited to strongly coupled systems where the ST states comprise both the ground and excited valley-orbit states~\cite{dehollain2014}. Furthermore, this method~\cite{dehollain2014} has limited fidelity as it relies on spin dependent tunnel rates that cannot be independently controlled. Here we utilise an alternative technique to perform single-shot readout of ST states across two coupled donor sites in a regime suitable for quantum computing applications. Importantly, there is no need for any capacitive difference between the charge sensor and the two donor sites, as our method does not utilise a direct (1,1)-(2,0) charge transfer signal. Instead, we utilise an energy selective readout technique relying on relaxation of the metastable triplet state in the (1,1) configuration when pulsed into the (2,0) region. The method has been previously demonstrated in a time-averaged fashion~\cite{Sachrajda2012,PhysRevB.92.125434,Harvey2015}, however we employ threshold discrimination analysis (c.f. single-spin readout~\cite{elzerman2004}) for \textit{single-shot} readout with fidelity greater than 98\%---close to fault tolerant thresholds for surface-code quantum computation~\cite{PhysRevA.86.032324}.

The device shown in Fig.~\ref{fig:device} was fabricated using scanning tunnelling microscope (STM) hydrogen lithography. The patterned donor sites, $L$, and $R$ consist of 2 and 1 phosphorus atoms respectively, determined by examining the size of the lithographic patches~\cite{buch2013,weber2014} and their charging energies (see Supplementary Material). Gates, $\{G_L,G_M,G_R\}$ control electron numbers at $L$ and $R$, whereas $G_{SET}$ is predominantly coupled to the single-electron transistor (SET) charge sensor. The SET is composed of approximately 1000 phosphorus atoms and is 19${\pm}$1~nm from $L$ and $R$, allowing for electron loading and unloading, see Fig.~\ref{fig:device}b. The SET is operated with a 2.5~mV source-drain bias and has a charging energy of ${\sim}5$~meV. Further details of the fabrication methods have been published previously~\cite{fuhrer2009}. All data herein was taken inside a dilution refrigerator at 100~mK (electron temperature $\sim$200~mK).

\begin{figure*}
\includegraphics[width=1\textwidth]{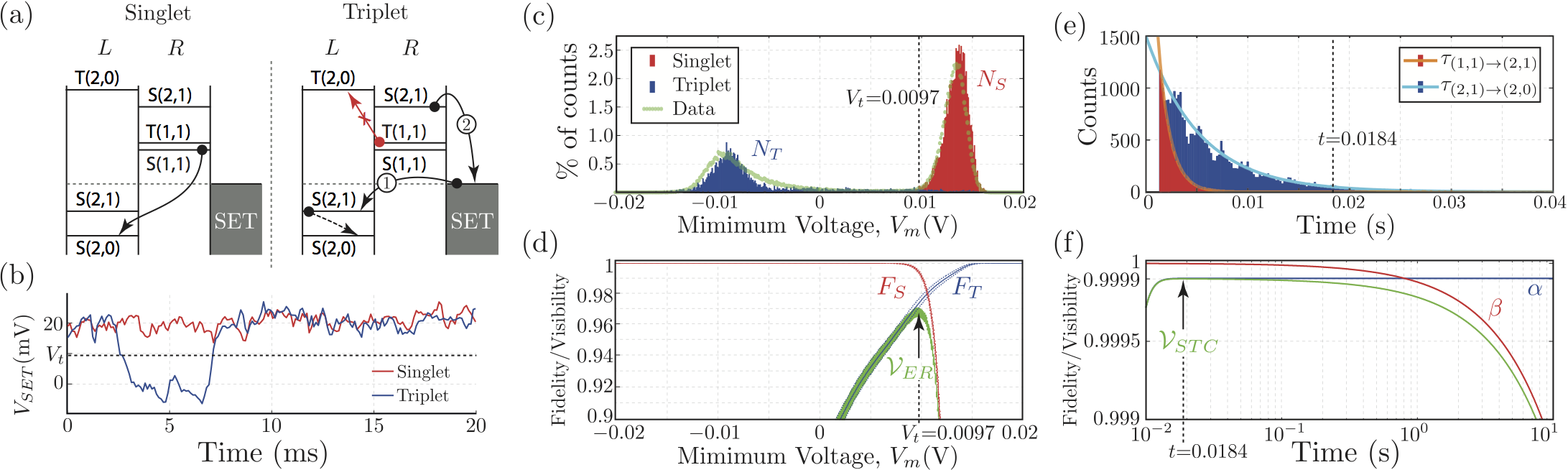}
\caption{{\bf Single-shot singlet-triplet readout in precision placed donor atoms.} a) The relevant chemical potentials of electrons at donor sites $L$ and $R$ with respect to the SET Fermi-level (grey region) at the readout position. The movement of electrons is shown by the solid black arrows and the red arrow depicts the forbidden transition of the T(1,1) state to the T(2,0) configuration due to Pauli spin blockade. b) Example SET readout trace of singlet and triplet states. (c,d) Optimisation of electrical readout visibility, $\mathcal{V}_{ER}$. Green markers in (c) show the minimum voltage after the current amplifier during readout for 100,000 traces. Solid bars show a simulation of 10,000 readout traces with a singlet (red) and triplet (blue) ratio of 1:2 as observed in the experiment. (d) The readout voltage threshold, $V_t$ is chosen to maximise $\mathcal{V}_{ER}$ (green) based on individual readout fidelities for singlet (blue), $F_S$, and triplet (red), $F_T$, states. Shaded regions of each line indicate one standard deviation. (e,f) Optimisation of state-to-charge conversion visibility, $\mathcal{V}_{STC}$. e) Experimentally obtained tunnel times (bars) for the relevant charge transitions and fits to exponential decays (lines). f) State-to-charge conversion fidelities for triplet and singlet, $\alpha$ (blue) and $\beta$ (red) respectively. Optimum readout time $\Delta t$ gives the maximum visibility $\mathcal{V}_{STC}$ (green). Analysis of ST readout was performed at $B_z{=}2.5$~T, where $T_1$ of the $\ket{{T^-}}$ state is of the order seconds.}
\label{fig:readout}
\end{figure*}

Figure~\ref{fig:device}c shows the SET current as a function of V$_{GL}$ and V$_{GR}$ near the (1,1)-(2,0) charge transition. No change in current is observed across the interdot transition (red line), and therefore ST readout cannot be performed here. Instead, we utilise two alternative tunnelling routes to the S(2,0) ground state near the (2,0)-(2,1) charge transition shown schematically in Fig.~\ref{fig:readout}a. Importantly, by monitoring the SET current in real time we can distinguish the two different tunnelling routes. At the readout position (white circle marker in Fig.~\ref{fig:device}c) the SET current is high when electrons are in the (2,0) and low in the (2,1) charge state. When initialising here in a S(1,1) state, an electron on $R$ can tunnel directly to $L$ forming S(2,0), because the SET is not sensitive to inter-donor transitions no charge transfer signal is observed (see left of Fig.~\ref{fig:readout}a). However, when initialising in any T(1,1) state at the readout position, tunnelling to T(2,0) is prohibited due to Pauli spin blockade~\cite{Ono1313,johnson2005,koppens2006,weber2014}. Now the S(2,0) ground state is reached via an electron first tunnelling \textit{onto} $L$ to form (2,1) (singlet state on left donor site, see arrow \circled{1} in Fig.~\ref{fig:readout}a) followed by an electron tunnelling from $R$ to the SET, forming the S(2,0), shown by arrow \circled{2} in Fig.~\ref{fig:readout}a. This process results in a `dip' in SET current~\cite{elzerman2004} which is used as the readout signal, see Fig.~\ref{fig:readout}b.

This charge transfer signal---given by movement of an electron to and from the SET---is as significant as for single-spin readout~\cite{buch2013}. Furthermore, the large ST energy splitting $\Delta_{ST}$ for donors---much larger than in gate-defined quantum dots---is reproducible as it is not influenced by surrounding electrostatic gates~\cite{weber2014}. The timescales for readout are dependent on the electron tunnel-on time from the SET to $L$, $\tau_{(1,1)\rightarrow(2,1)}$ (arrow \circled{1} in Fig.~\ref{fig:readout}a), and tunnel-off time from $R$ to the SET $\tau_{(2,1)\rightarrow(2,0)}$ (arrow \circled{2} in Fig.~\ref{fig:readout}a). These were determined by analysing $100,000$ readout traces to be $\tau_{(1,1)\rightarrow(2,1)}{=}1.15{\pm}0.03$~ms and $\tau_{(2,1){\rightarrow}(2,0)}{=}5.3{\pm}0.2$~ms. 

Following from previous works on single-shot spin readout~\cite{morello2010,pla2012,Veldhorst:2015qv,buch2013,watson2015}, the assignment of singlet or triplet state to each readout trace comprises two separate parts, (i) electrical readout and (ii) state-to-charge conversion (STC), which we discuss in detail below.

\textit{(i) Electrical readout.}---Here we determine whether a given SET current trace can be assigned as having a dip during the readout phase (time spent at the readout position), or not. In the experiment the SET current passes through a room temperature current amplifier, hence the resulting \textit{voltage} is relevant (see Fig.~\ref{fig:readout}b). During the readout phase a trace is assigned as having a voltage dip if its minimum value, $V_m{\le}V_t$. A Monte Carlo simulation of 10,000 readout traces with added white Gaussian noise equivalent to the experimental signal-to-noise ratio is shown in Fig.~\ref{fig:readout}c~\cite{morello2010}. This histogram shows the simulated minimum voltages $V_m$ from which we deduce the fidelity of assigning either a dip (triplet) or no dip (singlet), $F_T$ or $F_S$ respectively, to each trace using the equations,
\begin{align}
F_T &= 1-\int_{V_t}^{\infty}N_T(V_m)dV_m \\
F_S &= 1-\int_{-\infty}^{V_t}N_S(V_m)dV_m,
\end{align}
\noindent where $V_m$ is the minimum voltage and $N_i$ is the fraction of each state $i$. The results are shown in Fig.~\ref{fig:readout}d along with the calculated electrical readout visibility $\mathcal{V}_{ER}{=}F_T{+}F_S{-}1$. From Eqs.~1 and 2 we specify the optimum voltage threshold, $V_t$, where $\mathcal{V}_{ER}$ is maximised. In total, 500 independent simulations were run (each 10,000 simulated traces) allowing the assignment of errors shown in Table~\ref{tab}.

\textit{(ii) State-to-charge conversion.}---Next we determine the optimum \textit{readout time}, $\Delta t$, following the work on single-shot spin readout in~\cite{buch2013} and~\cite{watson2015}. The rate equation model described therein accounts for errors caused by, relaxation of excited states; triplet states failing to cause a tunnelling event before $\Delta t$; and a singlet state causing a tunnelling event within $\Delta t$. As inputs to the model, the tunnelling out time of the triplet state from $(1,1){\rightarrow}(2,1)$, $\tau_{T,\text{out}}$ is assigned the same value as $\tau_{(1,1)\rightarrow(2,1)}$, as shown in Fig.~\ref{fig:readout}e. The tunnel time, $\tau_{S,\text{out}}$, is also found experimentally by counting the number of tunnelling events occurring after a time much greater than $\tau_{T,\text{out}}$ (here we used $8$~ms) and attributing them to the exponential decay of the singlet state. Only $0.033\%$ of the 100,000 readout traces showed tunnelling after this time, giving an estimate of $\tau_{S,\text{out}}{=}16600\pm8300$s. Using these characteristic tunnelling times, we implement a rate equation model to determine the optimum readout time~\cite{buch2013}, $\Delta t$, based on the probability of successfully assigning a voltage dip a triplet or singlet state, $\alpha$ and $\beta$ respectively (see Supplementary Material). 

The resulting assignment probabilities $\alpha$ and $\beta$ are shown as a function of the readout time in Fig.~\ref{fig:readout}f. Similar to electrical readout, the visibility of state-to-charge conversion is calculated as $\mathcal{V}_{STC}{=}\alpha{+}\beta{-}1$, and the optimum readout time is chosen where $\mathcal{V}_{STC}$ is maximised and was found to be $\Delta t{=}18.4{\pm}0.7$~ms.

\begin{table}[h!]
\begin{center}
 \begin{tabular}{||c | c|| c | c||}
 \hline
 Elec. Readout & Value & STC Conv. & Value \\ [0.5ex]
 \hline\hline
 $V_t$~(V) & $0.0097{\pm}0.0007$ &   $\Delta t$~(ms) & $18.4{\pm}0.7$\\
 \hline
 $F_{S}$~$(\%)$ & $99.4{\pm}0.1$ &   $\alpha$~$(\%)$ & $99.990{\pm}0.001$\\
 \hline
 $F_{T}$~$(\%)$ & $97.3{\pm}0.3$ &  $\beta$~$(\%)$ & $99.999{\pm}0.001$\\
 \hline
$\mathcal{V}_{ER}$~$(\%)$ & $96.8{\pm}0.3$  &  $\mathcal{V}_{STC}$~$(\%)$ & $99.989{\pm}0.001$\\
 \hline
\end{tabular}
\end{center}
\vspace{-0.5cm}
\caption{{\bf Parameters for singlet-triplet readout.}}
\label{tab}
\end{table}

\begin{figure*}
\begin{center}
\includegraphics[width=1\textwidth]{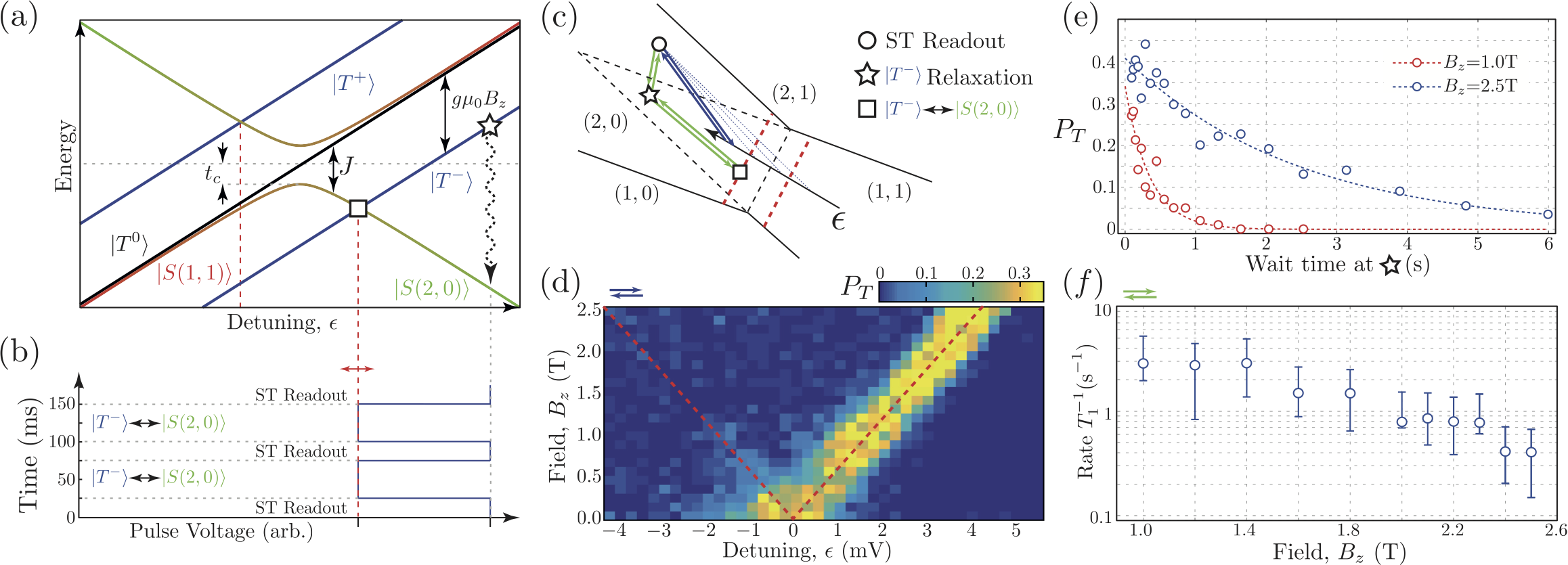}
\end{center}
\vspace{-0.5cm}
\caption{{\bf Field dependence of $\ket{S(2,0)}{\leftrightarrow}\ket{T^-}$ mixing and $\ket{T^-}$ $T_1$ relaxation.} a) The eigenspectrum of the two electron system at the (1,1)-(2,0) charge transition in a static magnetic field $B_z$. The detuning parameter $\epsilon$ corresponds to the black arrow shown in (c), and controls the exchange coupling, $J$, between the electrons. b) A schematic of the two-level pulse scheme used to observe mixing between $\ket{S(2,0)}{\leftrightarrow}\ket{T^-(1,1)}$ states. (c) Pulsing schemes for results shown in (d-f). d) The triplet probability, $P_T$, as a function of static magnetic field $B_z$ and the wait position along the detuning axis, $\epsilon$. A peak at $45^\circ$ for $\epsilon{>}0$ corresponds to the position in this parameter space where mixing between $\ket{S(2,0)}{\leftrightarrow}\ket{T^-(1,1)}$ can occur. A fainter peak running at $-45^\circ$ in the data corresponds to $\ket{S(2,0)}{\leftrightarrow}\ket{T^+(1,1)}$ mixing for $\epsilon{<}0$. (e,f) Measurement of the $\ket{T^-}$ $T_1$ time . To initialise the $\ket{T^-}$ state we wait at a position along $\epsilon$ shown by the square marker in (a,c) which follows the dotted red line in (d) for $\epsilon{>}0$ as a function of field, $B_z$. e) Probability of the triplet state for $B_z{=}\{1.0,2.5\}$~T, as a function of the time spent at the star marker shown in (c). f) Measurement of $T_1$ from $B_z{=}1.0{\rightarrow}2.5$~T.}
\label{fig:stvsfield}
\end{figure*}

Table~\ref{tab} gives a summary of the fidelity calculations, where the final measurement fidelity is given by, $F_M{=}\left(\alpha F_{T}{+}\beta F_{S}\right)/2{=}98.4{\pm}0.2\%$. Owing to the large energy separation between the S(2,0) and T(2,0) states, thermal broadening of the Fermi-distribution in the SET ($T_e{=}200$mK) has a negligible effect on the readout fidelity. As such, the $\mathcal{V}_{STC}$ is reported very close to unity. 

For this device the ST readout fidelity was limited by low electrical visibility $\mathcal{V}_{ER}$, itself restricted by a relatively low SET peak current of $30$~pA. Here we have a signal-to-noise ratio of 6.4, but with an increase of SET signal equivalent to $\text{\small{SNR}}{=}8$, we estimate achieving fidelities ${>}99$\%. Single-shot charge detection of this quality has previously been demonstrated in donor systems using DC biased SETs~\cite{watson2015,watson2015b} and with rf-reflectometry, which significantly improves SNR further still~\cite{house2016}. Nonetheless, high fidelity ST readout can be maintained over a large range of magnetic fields---as demonstrated in Fig.~\ref{fig:stvsfield}d---because $\Delta_{ST}$ is independent of $B_z$.

\textit{Singlet-triplet dynamics.}---As a demonstration of this readout technique, here we map out the $\ket{S(2,0)}{\leftrightarrow}\ket{T^-}$ anti-crossing as a function of $B_z$. Figure~\ref{fig:stvsfield}a shows the two electron eigenspectrum with the addition of the $\ket{S(2,0)}$ state as a function of detuning, $\epsilon$ (see arrow in Fig.~\ref{fig:device}c). To observe ST mixing we initialise deterministically in $\ket{S(2,0)}$ by performing ST readout. Next we apply a 50~ms pulse along the detuning axis toward the $\ket{S(2,0)}{\leftrightarrow}\ket{T^-}$ anti-crossing, allowing sufficient time for mixing (see Fig.~\ref{fig:stvsfield}b). Two of the triplet states $\{\ket{T^+}, \ket{T^-}\}$ are split by the Zeeman energy from the $\ket{S(1,1)}$ and $\ket{T^0}$ states, such that the position mixing between $\ket{S(2,0)}{\leftrightarrow}\{\ket{T^-},\ket{T^+}\}$ changes with $B_z$. The two-level pulse scheme is shown schematically by the blue arrows in Fig.~\ref{fig:stvsfield}c along with the ST mixing positions shown by the red dashed lines (not to scale). Finally, we pulse back to the ST readout position (circle marker) for 25~ms where we measure the triplet state probability, the results are shown in Fig.~\ref{fig:stvsfield}d. 

In addition to a clear $\ket{S(2,0)}{\leftrightarrow}\ket{T^-}$ mixing point for $\epsilon{>}0$, indicated by the high triplet probability in Fig.~\ref{fig:stvsfield}d (yellow), a faint feature related to $\ket{S(2,0)}{\leftrightarrow}\ket{T^+}$ mixing can also be seen at detuning values $\epsilon{<}0$. Mixing between $\ket{S(2,0)}{\leftrightarrow}\ket{T^+}$ is suppressed due to fast charge relaxation from $\ket{S(2,0)}{\rightarrow}\ket{S(1,1)}$ in this region. The position of $\ket{S(2,0)}{\leftrightarrow}\ket{T^-}$ mixing  in $\epsilon$ remains linear as a function of $B_z$, indicating a small value of tunnel coupling ($t_c$ in Fig.~\ref{fig:stvsfield}a), and hence no `spin-funnel' shape is seen as reported in similar experiments~\cite{maune2012,Veldhorst:2015qv}.

Finally, using the $\ket{S(2,0)}{\leftrightarrow}\ket{T^-}$ mixing point to randomly load the $\ket{T^-}$ state, we measure its $T_1$ lifetime using the three-level pulsing protocol shown by the green arrows in Fig.~\ref{fig:stvsfield}c. Relaxation of the $\ket{T^-}$ state occurs whilst inside the charge region enclosed by the dashed lines in Fig.~\ref{fig:stvsfield}c, i.e. where only \textit{inter donor-site} tunnelling is allowed. This position, indicated by the star marker in Fig.~\ref{fig:stvsfield}c, lies at $\epsilon{=}10~$mV, ensuring that $\ket{S(2,0)}$ remains the ground state for $B_z{\le}2.5$~T. The results for $T_1$ are shown in Fig.~\ref{fig:stvsfield}e and f. The observed decrease in $1/T_1$ as a function of increasing $B_z$ follows as a result of the decreasing energy gap between the excited $\ket{T^-}$ state and $\ket{S(2,0)}$ ground state~\cite{PhysRevB.82.241302}. Previous theoretical studies of triplet state relaxation in donors coupled along [001] predict a dependence on exchange energy as approximately, $1/T_1{\sim}J^3$~\cite{PhysRevB.82.241302}, and should be the focus of future experimental work.

High fidelity single-shot readout of individual and multiple qubit states is a prerequisite for the observation of post-classical multi-qubit phenomena, in particular two-qubit entanglement~\cite{Nielsen:2011}. In the original  Kane proposal for scalable donor based quantum computing architectures, the single-shot measurement of ST states is suggested to facilitate the readout of nuclear spins~\cite{kane1998silicon} and is advantageous over previously used readout techniques \cite{pla2013} as it does not require the high frequency manipulation of the electron spin. Furthermore, ST readout can be used to measure single electron spin qubits \cite{koppens2007} at lower magnetic fields and higher temperatures easing the constraints on microwave electronics and cryogenic cooling \cite{vandersypen2016}. Finally, encoding qubits using ST states \cite {kim2014quantum,laird2010,wu2015singlet} allows for an all electrical approach for control, in particular, multiple qubits can be coupled by utilising  the inherent electric dipole coupling given by the (1,1)-(2,0) charge configurations~\cite{Shulman202,Nichol:2017tg}. The results obtained herein, in addition to the reduced complexity of electron confinement in donors, makes a compelling case for further research on the scaling of multiple forms of donor based quantum computing architectures. 

\label{sec:disc}
\begin{acknowledgements}
This research was conducted by the Australian Research Council Centre of Excellence for Quantum Computation and Communication Technology (project no. CE110001027) and the US National Security Agency and US Army Research Office (contract no. W911NF-08-1-0527). M.Y.S. acknowledges an ARC Laureate Fellowship.
\end{acknowledgements}

\begin{widetext}

\vspace{10cm}

\setcounter{figure}{0}
\makeatletter 
\renewcommand{\thefigure}{S\@arabic\c@figure}
\makeatother

\section{\large{Supplemetary information}}

\section{Charging energy calculation}
\label{sec:energies}
In this section we provide a full derivation for the calculation of donor-island charging energies. From this analysis, along with examining the extent of the STM images, we can determine with high probability that the number of the donors at sites $L$ and $R$ are 2 and 1 respectively. 

The charging energy, of a quantum dot (QD) $a$, $\epsilon_c^a$, is calculated by establishing the charging energy of another QD in its vicinity, $b$ which is capacitively coupled to $a$~\cite{hile2015}. The charging energies of $a$ and $b$ are related through their mutual charging energy, $\epsilon_m$, given by,
\begin{equation}
\epsilon_m = \alpha_g^i \delta V_g^i,
\end{equation}
where $\alpha_g^i$ is the lever arm, or conversion factor, of voltage applied to gate $g$ to the energy of QD $i$ and $\delta V_g^i$ is the voltage shift of the potential of QD $i$ due to a charging event on QD $j$. Importantly, the mutual charging energy by definition must be equal for both QDs $a$ and $b$, that is to say we can write,
\begin{equation}
\alpha_g^a \delta V_g^a = \alpha_g^b \delta V_g^b.
\label{eq:mut}
\end{equation}
In a similar vein, the charging energy of a QD itself is given by the voltage difference of gate $g$ between two charge transitions, $\Delta V_g^i$, multiplied by its corresponding lever arm,
\begin{equation}
\epsilon_c^i = \alpha_g^i \Delta V_g^i.
\label{eq:chen}
\end{equation}
Now it is not necessary to know the level arms $\alpha_g^i$, as we can eliminate it by combining equations Eq.~\ref{eq:mut} and Eq.~\ref{eq:chen}, 
\begin{equation}
\frac{\epsilon_c^a}{\Delta V_g^a} \delta V_g^a = \frac{\epsilon_c^b}{\Delta V_g^b} \delta V_g^b.
\label{eq:equal}
\end{equation}

\noindent In the cases where multiple charging events occur within the range $\Delta V_g^i$, we may have the condition that $\epsilon_c^a \neq \epsilon_c^b$ and it is necessary to count the number of charging events. In this case, the measured $\Delta V_g^i$ is actually the sum of the true voltage change which we denote $\Delta \hat{V}_g^i$ and the number of charging events, $n^j$ of the other QD, that is,
\begin{equation}
\Delta V_g^i = \Delta \hat{V}_g^i + n^j \delta V_g^i,
\label{eq:hat}
\end{equation}
Substituting Eq.~\ref{eq:hat} into Eq.~\ref{eq:equal} we have the ratio of charging energies for the two QDs $a$ and $b$,
\begin{equation}
\frac{\epsilon_c^a}{\epsilon_c^b}  = \frac{\delta V_g^b(\Delta V_g^a - n^b \delta V_g^a)}{\delta V_g^a(\Delta V_g^b - n^a \delta V_g^b)}.
\label{eq:ratio}
\end{equation}

The above equation relates the charging energies of QDs $a$ and $b$. In our device, we use an SET island, which is a quantum dot with a small charging energy, such that we have the case $\Delta V_g^D{\gg}\Delta V_g^S$ (where we have denoted the donor-island as `$D$' and the SET as `$S$'). When the condition $\Delta V_g^D{\gg}\Delta V_g^S$ holds there will be multiple charging events of the SET i.e. $n^S{\neq}0$, and exactly zero for the donor ($n^D{=}0$) within the voltage ranges $\Delta V_g^D$ and $\Delta V_g^S$, respectively. As a result, we can simplify Eq.~\ref{eq:ratio} to,
\begin{equation}
\epsilon_c^{D} = \frac{\epsilon_c^{S}\delta V_g^{S}}{\Delta V_g^{S}} \Big(\frac{\Delta V_g^{D}}{\delta V_g^{D}} - n^{S}\Big).
\label{eq:final}
\end{equation}
Figure~\ref{fig:def} shows the measurement of all required parameters $\{\epsilon_c^S,\delta V_g^S, \delta V_g^D, \Delta V_g^S, \Delta V_g^D\}$, where we have chosen to measure along the right gate, $G_R$, i.e. $g{\rightarrow}R$. Using Eq.~\ref{eq:final} for the $1{\rightarrow}2$ electron transitions for both donor-islands, we find charging energies of $65{\pm}8$ and $43{\pm}5$ meV and for $L$ (2P) and $R$ (1P), respectively. These values are consistent with theoretical~\cite{saraiva2015,weber2014} and previously measured~\cite{fuechsle2012,weber2014} charging energies for 2P and 1P donor-islands respectively.

\section{Derivation of state-to-charge conversion}
Assuming we have detected a dip below the voltage threshold (see electrical readout section of main text), we now ask what is the probability that this event resulted from the presence of a T(1,1) state and not from the state S(1,1), causing a tunnelling event (1,1)${\rightarrow}$(2,1) (both scenarios would give rise to an equivalent voltage dip). This section was previously derived by Buch~\citep{Buch2013} for the purpose of single spin-readout, however, we repeat it here for completeness.

The calculated electrical readout fidelities $F_T$ and $F_S$ give indicate the fraction of $T$ and $S$ states at the \textit{end} of the readout time, but for quantum computation applications it is important to ascertain the loss in fidelity due to the change in these states during this time. This can be done by estimating the fraction of each state, $N^0_T$ and $N^0_S$ at the beginning of the readout phase. 

We define the initial time at the start of the readout phase as the elapsed time $t{=}0$, such that the system dynamics can be described by coupled rate equations in the basis $\{S,T\}$,
\begin{equation}
\frac{d \bar{N}(t)}{d t} = \begin{pmatrix}
-\frac{1}{\tau_{S,\text{out}}} & \frac{1}{T_1} \\
0 & -\frac{1}{\tau_{T,\text{out}}} - \frac{1}{T_1}
\end{pmatrix}\bar{N}(t),
\end{equation}
where $\tau_{S,\text{out}}$ and $\tau_{T,\text{out}}$ are the characteristic tunnelling out times of the singlet and triplet respectively. The relaxation time of the triplet $\ket{T^-}$ state, $T_1$ in the above equation, was measured at $B_z{=}2.5$T to be $T_1{=}3.0{\pm}0.3$~s (see Fig.~3e of main text). The above equation has the two solutions,
\begin{equation}
\bar{N}_T(t) = \bar{N}^0_T e^{-\frac{T_1 + \tau_{T,\text{out}}}{\tau_{T,\text{out}} T_1} t},
\end{equation}
\begin{align}
\bar{N}_S(t) = \frac{\bar{N}^0_S \tilde{T}^2 + \bar{N}^0_T \tau_{S,\text{out}} \tau_{T,\text{out}}}{\tilde{T}^2 e^{\frac{t}{\tau_{S,\text{out}}}}} - \frac{\bar{N}^0_T \tau_{S,\text{out}} \tau_{T,\text{out}}}{\tilde{T}^2 e^{-\frac{T_1 + \tau_{T,\text{out}}}{\tau_{T,\text{out}} T_1} t}},
\end{align}
where $\bar{N}_T(t)$ and $\bar{N}_S(t)$ are the remaining fraction of triplet and singlet states at time $t$ and ${\tilde{T}^2{=}T_1\tau_{S,\text{out}}{-}T_1\tau_{T,\text{out}}{+}\tau_{S,\text{out}}\tau_{T,\text{out}}}$, and $T_1$ is the characteristic relaxation time of the excited triplet state. Importantly, here we have used the relaxation of the triplet-minus state as described in the main text.

\begin{figure}
\begin{center}
\includegraphics[width=1\columnwidth]{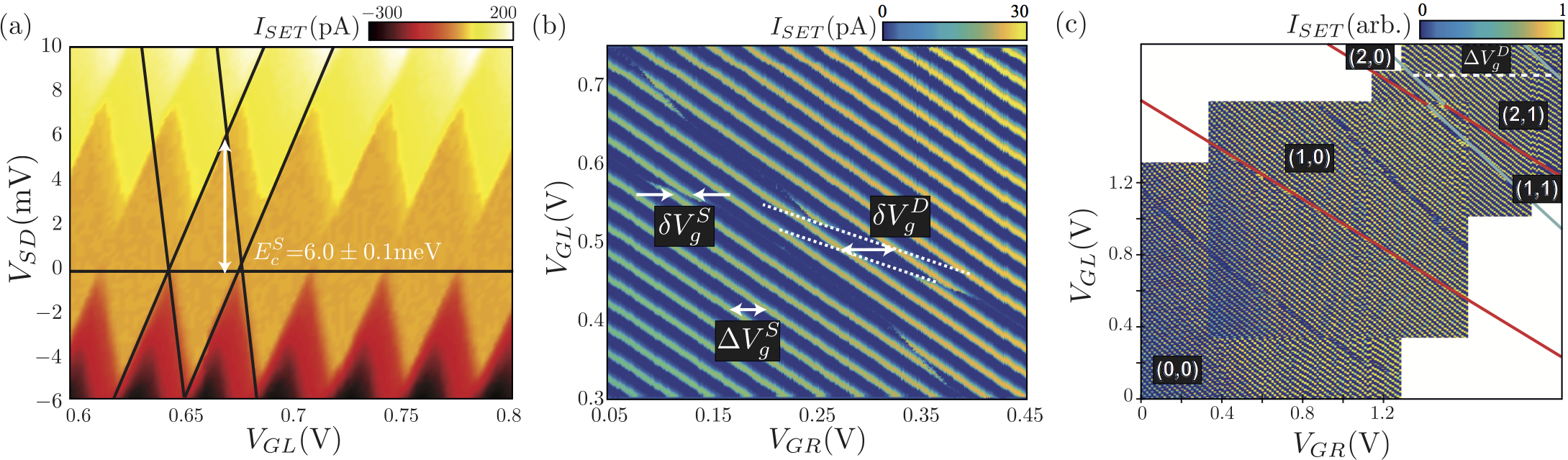}
\end{center}
\caption{{\bf Calculations of charging energy for the donor-islands $L$ and $R$.} a) The Coulomb diamonds used to measure the charging energy of the SET, $\epsilon_c^{S}$. b) A charge stability map ($V_{GL}$ vs. $V_{GR}$) showing the definition of the voltage parameters, $\Delta V_g^S$, $\delta V_g^S$ and $\delta V_g^S$. c) A composite charge stability map showing all the observed charge transitions in the device as a function of $V_{GL}$ and $V_{GR}$. The four separate maps vary in the middle gate voltage, $V_{GM}$, from $0.1{-}0.7$~V. Red and blue lines show charge transitions of donor-islands $L$ and $R$ respectively. Other observed transitions are attributed to charge traps in the vicinity of the SET and are not relevant to the experiment. The definition of $\Delta V_g^D$ is shown by the white dashed line. The number of SET charge transitions, $n^S$ are counted along this line.}
\label{fig:def}
\end{figure}

Next we define the fraction of states that successfully generate a voltage dip by time $t$ as, 
\begin{equation}
N^{'}(t) = 1 - \bar{N}_T(t) - \bar{N}_S(t) = \alpha(t) \bar{N}^0_T + (1 - \beta(t)) \bar{N}^0_S
\end{equation}
equivalently, the faction of states that do not generate a voltage dip at time $t$ is given by $1-N^{'}(t)$, or,
\begin{equation}
1-N^{'}(t) = \bar{N}_T(t) + \bar{N}_S(t) = (1 - \beta(t)) \bar{N}^0_T + \alpha(t) \bar{N}^0_S.
\end{equation}
\noindent Here, $\alpha$ represents the probability of the triplet state tunnelling to the SET from the double-donor system (with correction for its $T_1$ relaxation), while $\beta$ represents the probability of the singlet state \textit{not} tunnelling to the SET i.e. at $t{=}0$ the probability the singlet state has not tunnelled to the SET is unity. 

From the above set of relationships we can calculate the probability that the detected dip at time $t$ can be assigned to a $T$ or $S$ state. We denote these as $\alpha(t)$ and $\beta(t)$, respectively,
\begin{equation}
\alpha(t){=}\frac{1}{\tilde{T}^2}[\tilde{T}^2{-}\frac{\tau_{S,\text{out}}\tau_{T,\text{out}}}{e^{\frac{t}{\tau_{S,\text{out}}}}}{-}\frac{T_1\tau_{S,\text{out}}}{e^{\frac{t}{\tau_{T,\text{out}}}}e^{\frac{t}{T_1}}}{+}\frac{T_1\tau_{T,\text{out}}}{e^{\frac{t}{\tau_{T,\text{out}}}}e^{\frac{t}{T_1}}}],
\end{equation}
\begin{equation*}
\beta(t) = e^{-\frac{t}{\tau_{S,\text{out}}}}.
\end{equation*}
Now, to optimise the state-to-charge conversion fidelity we find the maximum visibility given by,
\begin{equation}
\mathcal{V}_{STC}{=}\alpha(t){+}\beta(t){-}1.
\end{equation}

\end{widetext}

\end{document}